\documentclass[aps,prl,reprint,superscriptaddress]{revtex4-1}

\usepackage[utf8]{inputenc}	

\usepackage{hyperref} 
\hypersetup{ 
     colorlinks=true, 		% colorise les liens 
     urlcolor= Blue3,  		% couleur des hyperliens (doit inclure x11names dans xcolor ci-dessus)
     linkcolor= Blue3, 		% couleur des liens internes 
     citecolor=Green4,		% couleur des liens de citation
}
\usepackage{amssymb} 		% ajoute des symboles mathématiques
\usepackage{amsfonts}		% ajoute des polices mathématique
\usepackage{amsmath}        % ajoute des environnements mathématiques
\usepackage{bm}				% ajoute des caractères grecs en gras
\usepackage{mathrsfs}		% ajoute une meilleure police calligraphique pour certains symboles
\usepackage{graphicx}		% gère l'insertion des figures
\usepackage[x11names,svgnames]{xcolor}       % gère plus de couleurs
\usepackage{color}		
\usepackage{float}
\newcommand{\cmmnt}[1]{}
\begin{document}

\title{Photon-assisted Dynamical Coulomb Blockade in a Tunnel Junction}
\author{Samuel Houle}
\email{Samuel.Houle@USherbrooke.ca}
\affiliation{Département de physique, Université de Sherbrooke, Sherbrooke, Québec, J1K 2R1, Canada}
\author{Karl Thibault}
\email{Karl.Thibault@usherbrooke.ca}
\affiliation{Département de physique, Université de Sherbrooke, Sherbrooke, Québec, J1K 2R1, Canada}
\author{Edouard Pinsolle}
\email{Edouard.Pinsolle@USherbrooke.ca}
\affiliation{Département de physique, Université de Sherbrooke, Sherbrooke, Québec, J1K 2R1, Canada}
\author{Christian Lupien}
\email{Christian.Lupien@USherbrooke.ca}
\affiliation{Département de physique, Université de Sherbrooke, Sherbrooke, Québec, J1K 2R1, Canada}
\author{Bertrand Reulet}
\email{Bertrand.Reulet@USherbrooke.ca}
\affiliation{Département de physique, Université de Sherbrooke, Sherbrooke, Québec, J1K 2R1, Canada}
\date{\today}
%=====================================================
\begin{abstract}
We report measurements of photon-assisted transport and noise in a tunnel junction in the regime of dynamical Coulomb blockade. We have measured both dc non-linear transport and low frequency noise in the presence of an ac excitation at frequencies up to 33 GHz. In both experiments, at very low temperatures, we observe replicas at finite voltage of the zero bias features, a phenomenon characteristic of photon emission/absorption. However, the ac voltage necessary to explain our data is notably different for transport and noise, indicating that usual theory of photon-assisted phenomena fails to account for our observations. 
\end{abstract}
\maketitle

%=====================================================
\emph{Introduction}. The physics of Dynamical Coulomb Blockade (DCB) in a tunnel junction - two metallic contacts separated by a thin insulating barrier - can be understood in simple terms: when an electron tunnels through the barrier, the charge on the contacts suddenly changes by an amount equal to $e$. This results in a sudden change in the voltage $V$ across the junction given by $e/C$ with $C$ the capacitance between the contacts. Since the statistics of the electron tunneling across the barrier is driven by $V$ , it follows that next electrons have less probability to cross the barrier\cite{Levitov1996}. When two tunnel junctions are in series, thus defining a conducting island between the two insulating barriers, a complete blocking of electron transport may occur at very low temperature $k_BT\ll e^2/C$, referred to as Coulomb blockade \cite{Grabert1992}. With one junction only, the charge accumulated in the contacts leaks continuously into the rest of the circuit, leading to a progressive recovery of the possibility to have subsequent tunneling events \cite{Averin1992}. This phenomenon occurs on a time scale that depends on the circuit in which the junction is embedded, given for example by the product $RC$ if the electromagnetic environment of the junction consists of a resistor $R$. This mechanism gives its dynamical aspect to the DCB, which can be seen as a frequency-dependent feedback of the electromagnetic environment of the junction on electron transport. It is the same feedback that leads to corrections to high order cumulants of current fluctuations (in both cases current fluctuations generated by the environment also matter) \cite{Beenakker2003, Reulet2003, Reulet2005}.

The theory of DCB is well established \cite{Devoret1990,Ingold1992} and has been successfully confirmed by many transport experiments \cite{Delsing1989,Geerligs1989,Cleland1992,Holst1994,Pierre2001}. It consists of describing the electromagnetic environment of the junction by a collection of modes characterized by their density $P(E)$ and population. In this picture, the environment authorizes inelastic tunneling of the electrons, which is possible only if the energy gained or lost by the electron can be accommodated by the environment. While very efficient, this description however seems to rely on different ingredients than the description of DCB by feedback. In particular, the dynamical response of the environment does not appear in a clear way.

Here we report an experiment that aims at probing the dynamical aspect of DCB by measuring the dc transport in a tunnel junction embedded in a resistive environment (the standard configuration for the observation of DCB) in the presence of an ac voltage. The frequency $f$ and amplitude $V_{ac}$ of the excitation can be tuned to explore different regimes, from weak $eV_{ac}\ll hf$ to strong $eV_{ac}\gg hf$, and from slow $f<(RC)^{-1}$ to fast $f>(RC)^{-1}$ excitation. Together with the dc nonlinear transport, we measure the zero frequency current noise to determine the ac voltage experienced by the junction, thus leaving no free parameter to fit the experimental transport data. We show that our transport data cannot be explained by recent theories that include the periodic voltage drive into the $P(E)$ theory \cite{Safi2010,Grabert2015} if we use the values of the ac voltage given by the noise measurements.

\begin{figure}[t!]\centering
\includegraphics[width=0.9\columnwidth]{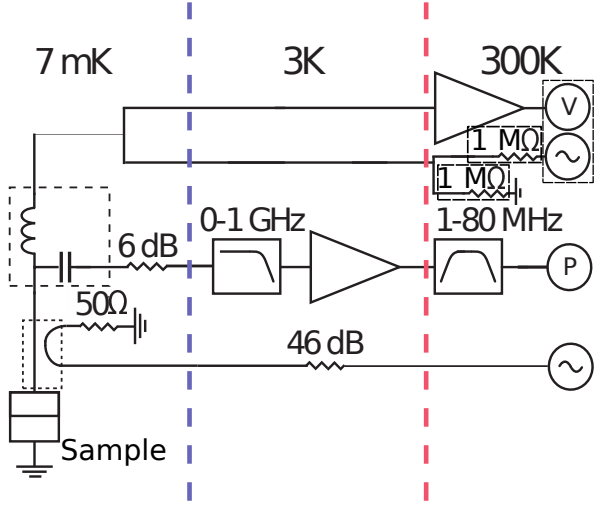}
\caption[]{(color online) Experimental setup. The (V) symbol represents a voltmeter / lock-in. The (P) symbol represents an RF power detector.}
\label{fig:setup}
\end{figure}

\emph{Experimental Setup}. Measurements are carried out on a 1~$\mu$m $\times$ 500~nm Al/Al oxide/Al tunnel junction  fabricated by electrolitography using the Dolan bridge technique\cite{Dolan1977}. The sample is placed in a dilution refrigerator of base temperature $8$~mK. A strong permanent magnet placed underneath the sample keeps the aluminum in its normal, non-superconducting state. The experimental setup is shown in Fig.~\ref{fig:setup}.
The junction is dc biased through a bias tee and ac biased through a 16~dB directional coupler placed between the bias tee and the junction. The dc biasing is performed through a 1~M$\Omega$ resistor at room temperature and cryogenic low pass filters (not shown). Thus the sample of resistance $R_\infty=620$~$\Omega$ (measured at high voltage), is effectively current biased at low frequency. We measure both the dc voltage $V$ across the junction and the differential resistance $\mathrm{d}V/\mathrm{d}I$ with usual lock-in technique%, with an excitation current of 1~nA at 77~Hz%
. The sample's current fluctuations are attenuated by a 6~dB cryogenic attenuator placed at the lowest temperature, amplified in the range 1~MHz-1~GHz by a cryogenic amplifier on the 3~K stage of the refrigerator, then further amplified at room temperature and bandpass filtered in the 1-80~MHz range. The power $P$ of the resulting signal is measured by a power detector as a function of the dc current $I$ as well as its derivative $\mathrm{d}P/\mathrm{d}I$. The purpose of the attenuator is threefold. First it offers the junction a well defined resistive environment so that the electromagnetic environment of the junction is well described by its geometric capacitance $C$ in parallel with a resistor $R_0=50~\Omega$. Second, it attenuates the noise emitted by the amplifier towards the junction by a factor $\alpha=10^{-0.6}\simeq0.25$, thus keeping the electron temperature reasonably low (according to our noise measurements, $T=50$~mK, see below). Third, it attenuates the noise coming from the amplifier and reflected by the sample. At low frequency, the impedance of the sample is purely resistive, so the low frequency  reflection coefficient is $\Gamma=(R-R_0)/(R+R_0)$, with $R$ the differential resistance of the junction ($\Gamma\simeq 0.85$ at high voltage). The detected power $P$ is related to the noise spectral density $S$ of the tunnel junction by:
\begin{equation}
P=G\left[ \alpha(1-\Gamma^2) T_N + \alpha^2\Gamma^2 T_{A1} + T_{A2} \right]
\label{eq:P}
\end{equation}

\noindent where $T_N=RS/(2k_B)$ is the noise temperature of the sample, $T_{A1}$ the equivalent temperature of the noise emitted by the amplifier towards the sample, $T_{A2}$ the equivalent temperature of the noise added by the setup and $G$ the power gain of the setup. Since $R$ depends on the bias, so does $\Gamma$. Thus the bias dependence of $P$ comes from that of the noise generated by the sample as well as from the noise of the amplifier being reflected in a bias-dependent way. Since $T_{A1}\gg T_N$, the presence of the attenuator is essential. The amount of attenuation is chosen so as to keep enough signal while greatly attenuating the reflection of the amplifier's noise. 
An ac excitation can be sent to the sample through the 2-40 GHz coupler placed next to it. The relevant frequency scale $f_c$ for the circuit is given by an inverse "RC" time where "R" corresponds to the parallel combination of $R$ and $R_0$, i.e. $f_c\simeq(2\pi R_0C)^{-1}$. In order to know $C$ we have measured the frequency dependence of both the reflection coefficient of the bare junction and the noise emitted by the junction with high bias (data not shown). We observe $f_c\simeq 14$~GHz.

\begin{figure}
\includegraphics[width=\columnwidth]{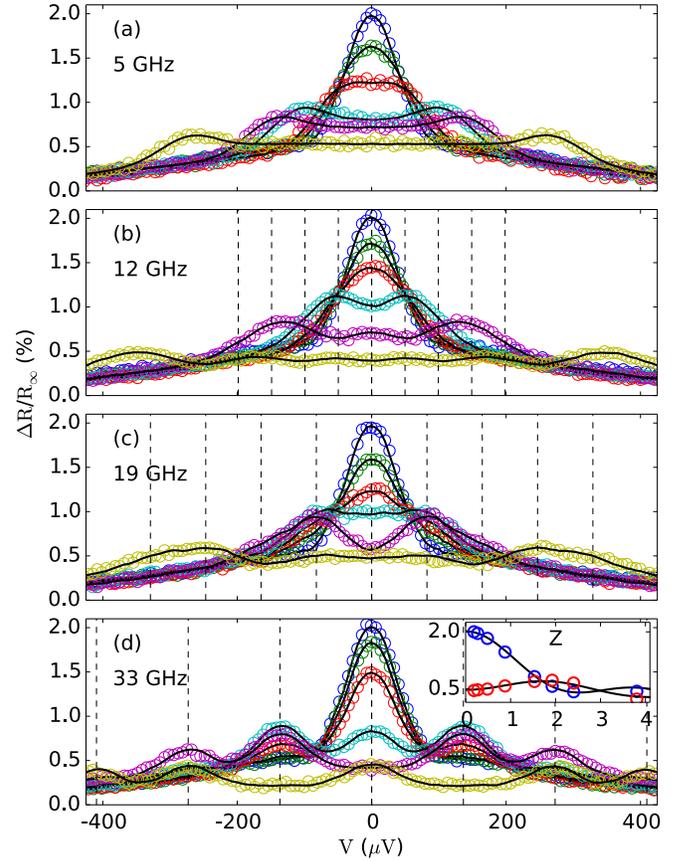}
\caption[]{(color online) Rescaled differential resistance vs. bias voltage. (a): $f=5$~GHz (b): $f=12$~GHz, (c): $f=19.9$~GHz, (d): $f=33$~GHz. In all graphs, symbols are experimental data and black, solid lines fits using theory of Eq. (\ref{eq:Tien}). The blue symbols correspond to $V_{ac}=0$. The different colors represent different ac excitation amplitudes $z=eV_{ac}/hf$: (a): $z=0$, $2.05$, $3.67$, $6.53$, $8.22$, $14.62$; (b): $z=0$, $0.87$, $1.24$, $1.96$, $3.90$, $9.40$; (c): $z=0$, 0.8, 1.14, 2.05, 2.88, 4.58; (d): $z=0$, 0.48, 0.88, 1.53, 2.40, 3.81. The dashed vertical lines correspond to $V=nhf/e$ with $n$ integer. Inset of (d): $\Delta R/R_\infty$ measured at $V=0$ (blue) and  $V=hf/e$ (red) as a function of $z$ at $f = 33$~GHz.}

\label{fig:resistance}
\end{figure}

\emph{Results: differential resistance.} We show in Fig.~\ref{fig:resistance} the measurement of $\Delta R/R_{\infty}$ vs. $V$ at very low electron temperature ($T=50$~mK), with $\Delta R=\mathrm{d}V/\mathrm{d}I-R_{\infty}$ the excess resistance. In the absence of ac excitation (dark blue curves in Fig.~\ref{fig:resistance}(a)-(d)) we observe a peak at zero bias, characteristic of DCB. This peak decreases when $T$ increases and disappears at $T\sim0.6$~K. In the presence of ac bias of frequency $f$, we observe replicas of the central DCB peak that appear at finite voltage. Fig.~\ref{fig:resistance} shows our results for excitations $f=5$~GHz (a), 12~GHz (b), 19.9~GHz (c) and 33~GHz (d). The positions of the replicas depend on the excitation frequency $f$ and are given by $V=nhf/e$ with $n$ integer (vertical dashed lines in Fig.~\ref{fig:resistance}). At low excitation frequency $f=5$~GHz, there is a strong overlap between the replicas. As the frequency increases, the peaks are more and more separated, and clearly distinguishable at $f=19$~GHz and $f=33$~GHz. We observe that the amplitude of the various peaks, including the one at zero voltage, depends in a non-trivial way on ac voltage, clearly indicating that our observations are not due to simple heating of the electrons by the ac excitation, see inset of Fig.~\ref{fig:resistance}(d).

\emph{Interpretation: Tien-Gordon mechanism.} Similar observations have been performed a long time ago with Josephson junctions \cite{Tien1963}. These observations are well understood if the only effect of the ac excitation is to induce a time-dependent, homogeneous voltage in one of the contacts. This voltage induces a time dependent phase in the electron wavefunctions in the contact given by $\varphi(t)=\int eV(t')/\hbar \;\mathrm{d}t'$. For a sinewave excitation $V(t)=V+V_{ac}\cos2\pi ft$, this phase results in the $I(V)$ (current-voltage) and $S(V)$ (noise-voltage) characteristics of the Josephson junction being modified into:
\begin{align}
\begin{split}
I(V)&=\sum_{n=-\infty}^{\infty}J_n^2(z) I_0(V+nhf/e)\\
S(V)&=\sum_{n=-\infty}^{\infty}J_n^2(z) S_0(V+nhf/e)
\end{split}
\label{eq:Tien}
\end{align}
with $I_0(V)$ and $S_0(V)=eI_0(V)\mathrm{coth}(eV/2k_BT)$ the current and noise characteristics of the junction without ac excitation. Here $J_n$ are the Bessel functions of the first kind and $z=eV_{ac}/hf$. This formula predicts that the $I(V)$ characteristics of the junction is a superposition of many copies of the original characteristic shifted in voltage by $nhf/e$ with weights given by $J_n^2(z)$. Clearly this result is not restricted to the $I(V)$ characteristic of Josephson junctions but should apply to a broader range of quantities and devices. An important condition is however the absence of dynamical degrees of freedom in the system. Let us consider for example a mesoscopic diffusive wire. Since such a conductor is linear, its $I(V)$ characteristics is not modified by the ac excitation, in agreement with Eq.~(\ref{eq:Tien}), since $\sum J_n^2(z)=1$. However, its shot noise does show replicas that are well described by Eq.~(\ref{eq:Tien}) \emph{provided the excitation frequency is low when compared to the inverse diffusion time of the wire}. Such replicas have been observed experimentally in the noise of wires \cite{Schoelkopf1998}, tunnel junctions \cite{Gabelli2008} and quantum point contacts \cite{Zakka2007}. The breakdown of Eq.~(\ref{eq:Tien}) at high frequencies has been calculated \cite{Shytov2005, Bagrets2007} but not yet observed.
The condititions of validity of Eq.~(\ref{eq:Tien}) can also be understood in the Landauer-Büttiker framework of quantum transport, where the transport properties of the sample are embedded in transmission coefficients. In this description, Eq.~(\ref{eq:Tien}) is obtained only if the transmission coefficients are energy independent, as for a tunnel junction. However, DCB induces energy dependence in the transmissions even in a tunnel junction \cite{Matveev1993,Kindermann2003}, thus Eq.~(\ref{eq:Tien}) might not hold. Furthermore, following \cite{Kindermann2003}, at low voltage and temperature, the strength of the energy dependence should be influenced by the environmental impedance at the excitation frequency $f$, thus DCB might be different for $f>f_c$ or $f<f_c$. Our experiment allows to probe the two regimes, where the charge has time to be evacuated before the ac voltage reverses or not. Recent theoretical papers have explicitly addressed the case of DCB in a tunnel junction in the presence of an ac voltage \cite{Safi2010,Grabert2015}. Despite what we just discussed, the prediction is that Eq.~(\ref{eq:Tien}) should be still valid. The only dynamical effect is that the ac voltage seen by the sample depends on frequency since the electromagnetic environment acts as a low-pass filter for the excitation. It is thus crucial to have a way to know the actual $V_{ac}$ experienced by the sample, which is provided by the measurement of low frequency noise.

\begin{figure}[t!]\centering
\includegraphics[width=\columnwidth]{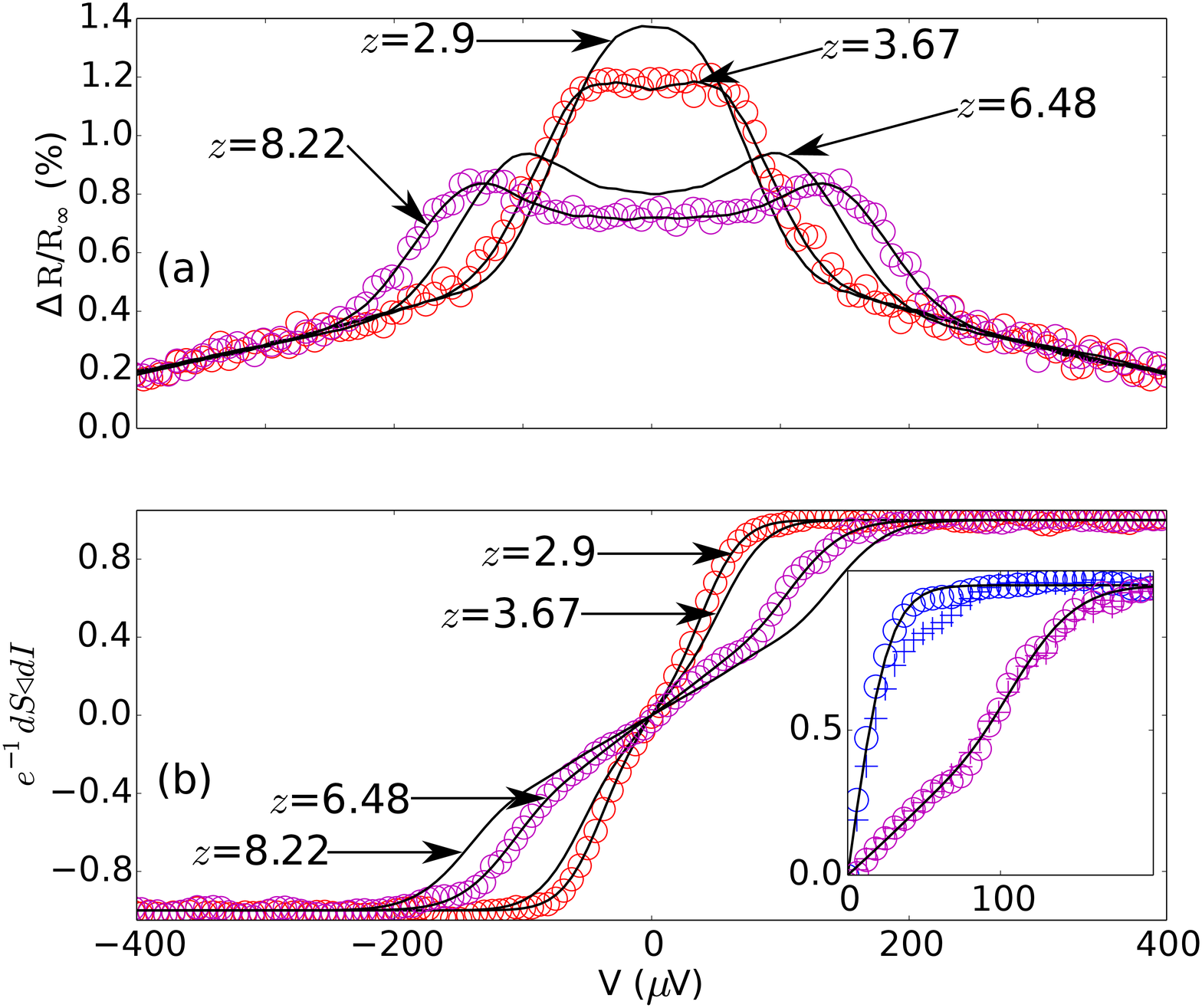}
\caption[]{(color online) Normalized differential resistance (a) and noise (b) vs. bias voltage for an ac excitation at 5 GHz. Symbols are experimental data and black lines are fits. Curves at $z=2.9$ and $z=6.48$ are obtained by fitting noise; those at $z=3.67$ and $z=8.22$ by fitting transport. Inset of (b): '+' symbols are experimental raw data, circles are data after subtraction of the noise emitted by the amplifier being reflected by the sample. Black lines are fits of the corrected data using $S=eI\mathrm{coth}(eV/2k_BT)$ for $z=0$. Blue: $z=0$. Purple: finite ac bias, well fitted by $z=6.48$.}
\label{fig:Diff}
\end{figure}

\emph{Results: noise}. In order to probe potential dynamical effects in the photon-assisted DCB, it is necessary to calibrate the voltage experienced by the junction for each excitation frequency. We have done this by measuring the low frequency photon-assisted noise of the junction as in \cite{Gasse2013}. We show in Fig.~\ref{fig:Diff}(b) results for the normalized differential noise $e^{-1}\mathrm{d}S/\mathrm{d}I$, for an ac excitation at $f=5$~GHz. DCB corrections to the noise are of the order of a few \% and can be neglected \cite{Altimiras2014}, so we can fit our data with the usual theory \cite{Levitov1994}. However corrections due to the bias-dependent reflection of the noise of the amplifier must be taken into account, according to Eq.~(\ref{eq:P}). The only unknown parameter is $T_{A1}$, the noise of the amplifier that is emitted towards the sample, which we determine as follows. The differential noise vs. $V$ in the absence of ac excitations should look like a step function, the width of which is inversely proportional to the temperature. As shown in the inset of Fig.~\ref{fig:Diff}(b), the experimental curves show additional steps. These steps are very well taken into account if we fit the data by Eq.~(\ref{eq:P}) with $T_{A1}=3$~K \footnote{Similar extra steps have been reported in \cite{Altimiras2014} and attributed to the noise being measured given by non-symmetrized current-current correlator. This discussion is irrelevant here since our measurement corresponds to zero frequency noise, for which commutativity of current operators is not an issue.}. This represents about half the total noise temperature of the amplifier, which is typical for such component. We show in Fig.~\ref{fig:Diff}(b) the normalized differential noise after correction. Corrected data are very well fitted by usual theory (Eq.~(\ref{eq:Tien})) with an electron temperature varying from $T=50$~mK at $V_{ac}=0$ up to $T=180$~mK at the highest excitations (33 GHz, $z\sim$~3.8). This apparent heating of the electrons by the microwave excitation does not appear in the transport data: photon-assisted transport is very well fitted by considering replicas of the data taken at $V_{ac}=0$, i.e. at $T=50$mK: peaks in $\Delta R/R_{\infty}$ are not broadened by the ac excitation. Taking replicas of transport measured at higher temperature yields to a much poorer fit. From the noise measurements we can infer the renormalized ac voltage $z$ across the junction for each excitation frequency. Note that the contribution of the reflected noise is most important when $\Gamma$ depends strongly on bias, i.e. mainly at low ac excitation: in the inset of Fig. \ref{fig:Diff}(b) while the raw data at $z=0$ (blue crosses) clearly have an extra step as compared with the corrected data (blue circles), the effect of the correction becomes negligible at $z=6.48$ (purple crosses/circles).

\begin{figure}[t!]\centering
\includegraphics[width=1.05\columnwidth]{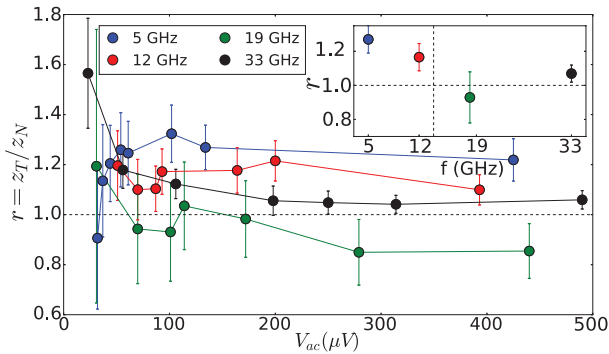}
\caption[]{Ratio $r=z_T/z_N$ between the normalized ac voltages $z=eV_{ac}/hf$ used to fit transport data ($z_T$) and noise ($z_N$) as a function of the ac voltages obtained from noise. Inset: ratio $r$ as a function of frequency. The values assigned to $r$ for each frequency are obtained by averaging over $r$ for every value of $V_{ac}$ above 50 $\mu V$ for that particular frequency. The vertical dashed line indicates the characteristic frequency $f_c=14$~GHz.}
\label{fig:Z3}
\end{figure}

\emph{Interpretation: effective ac voltage}.
In order to compare experiment with theory we fit the transport data using Eq.~(\ref{eq:Tien}) leaving the ac voltage as the only fitting parameter (solid lines on Fig.~\ref{fig:resistance}). For the $I_0(V)$ characteristics in the absence of ac excitation, we take the experimental data at $z=0$. The agreement between theory and data is excellent. Then we compare the rescaled values of the ac voltage used to fit the transport data $z_T$ with those extracted from our noise measurements $z_N$: we show in Fig.~\ref{fig:Z3} the ratio $r=z_T/z_N$ as a function the ac excitation obtained from noise measurements. While for low excitation the uncertainty on $z$ does not allow for a definitive conclusion, one clearly observes $r \neq 1$ at high ac voltages. In particular, all values of $r$ are greater than unity for $V_{ac}>50 \mu V$ at all frequencies except $19$ GHz for which $r\sim1$. An example of the discrepancy in $z$ between transport and noise is reported in Fig.~\ref{fig:Diff} where differential resistance (a) and differential noise (b) are plotted vs. dc voltage for two excitation powers at $f=5$~GHz. While each of the curves can be very well fitted, it is clearly impossible to fit both transport and noise with the same $V_{ac}$. The ratio $r$ is plotted as a function of frequency in the inset of Fig. \ref{fig:Z3}. The characteristic frequency $f_c=14$~GHz is indicated as a vertical dashed line. One should also have $r=1$ at low enough frequency.

A measurement of photo-assisted transport in the DCB regime has been reported in \cite{Parlavecchio2015}. Unfortunately, only one frequency and one value of the ac voltage ($z=1.15$) have been measured, and there was no way to calibrate $z$ beside estimating the attenuation in the circuit. Thus the phenomenon we report here could not have been observed in \cite{Parlavecchio2015}.

\emph{Conclusion.} Our experiments show that Tien-Gordon mechanism applied to the usual theory of transport and noise of a tunnel junction in a resistive environment does not explain our observations, suggesting the existence of dynamical phenomena not well taken into account. For example, the ac voltage experienced by the sample should depend on the value of the ac complex impedance of the junction at the excitation frequency, which itself depends of the dc and ac bias of the sample \cite{Joyez1998, Parlavecchio2015, Altimiras2016}. The ac excitation of the sample may also induce non-thermal distribution functions in the contacts that influence differently the noise and transport. \footnote{Indeed, we note that our photon-assisted transport data are well fitted by taking replicas of the data at $V_{ac}=0$ and the lowest temperature, whereas noise data at high excitation show an increased effective temperature.}\footnote{Such effects have been discussed in similar samples where the amplitude of quantum oscillations could not be explained by the ac voltage excitation deduced from photon-assisted noise measurements \cite{Gasse2013}.} More experiments are called for, for example with a different electromagnetic environment, or using non-classical light as recently proposed \cite{Souquet2014}, in order to probe more in depth the dynamical properties of DCB.

We acknowledge fruitful discussions with A. Clerk. We thank G. Laliberté for technical help. This work was supported by the Canada Excellence Research Chairs, Government of Canada, Natural Sciences and Engineering Research Council of Canada, Québec MEIE, Québec FRQNT via INTRIQ, Université de Sherbrooke via EPIQ, and Canada Foundation for Innovation.
\bibliography{PA_DCB_v_2}

\begin{thebibliography}{34}%
\makeatletter
\providecommand \@ifxundefined [1]{%
 \@ifx{#1\undefined}
}%
\providecommand \@ifnum [1]{%
 \ifnum #1\expandafter \@firstoftwo
 \else \expandafter \@secondoftwo
 \fi
}%
\providecommand \@ifx [1]{%
 \ifx #1\expandafter \@firstoftwo
 \else \expandafter \@secondoftwo
 \fi
}%
\providecommand \natexlab [1]{#1}%
\providecommand \enquote  [1]{``#1''}%
\providecommand \bibnamefont  [1]{#1}%
\providecommand \bibfnamefont [1]{#1}%
\providecommand \citenamefont [1]{#1}%
\providecommand \href@noop [0]{\@secondoftwo}%
\providecommand \href [0]{\begingroup \@sanitize@url \@href}%
\providecommand \@href[1]{\@@startlink{#1}\@@href}%
\providecommand \@@href[1]{\endgroup#1\@@endlink}%
\providecommand \@sanitize@url [0]{\catcode `\\12\catcode `\$12\catcode
  `\&12\catcode `\#12\catcode `\^12\catcode `\_12\catcode `\%12\relax}%
\providecommand \@@startlink[1]{}%
\providecommand \@@endlink[0]{}%
\providecommand \url  [0]{\begingroup\@sanitize@url \@url }%
\providecommand \@url [1]{\endgroup\@href {#1}{\urlprefix }}%
\providecommand \urlprefix  [0]{URL }%
\providecommand \Eprint [0]{\href }%
\providecommand \doibase [0]{http://dx.doi.org/}%
\providecommand \selectlanguage [0]{\@gobble}%
\providecommand \bibinfo  [0]{\@secondoftwo}%
\providecommand \bibfield  [0]{\@secondoftwo}%
\providecommand \translation [1]{[#1]}%
\providecommand \BibitemOpen [0]{}%
\providecommand \bibitemStop [0]{}%
\providecommand \bibitemNoStop [0]{.\EOS\space}%
\providecommand \EOS [0]{\spacefactor3000\relax}%
\providecommand \BibitemShut  [1]{\csname bibitem#1\endcsname}%
\let\auto@bib@innerbib\@empty
%</preamble>
\bibitem [{\citenamefont {Levitov}\ \emph {et~al.}(1996)\citenamefont
  {Levitov}, \citenamefont {Lee},\ and\ \citenamefont {Lesovik}}]{Levitov1996}%
  \BibitemOpen
  \bibfield  {author} {\bibinfo {author} {\bibfnamefont {L.~S.}\ \bibnamefont
  {Levitov}}, \bibinfo {author} {\bibfnamefont {H.}~\bibnamefont {Lee}}, \ and\
  \bibinfo {author} {\bibfnamefont {G.~B.}\ \bibnamefont {Lesovik}},\ }\href
  {\doibase 10.1063/1.531672} {\bibfield  {journal} {\bibinfo  {journal}
  {Journal of Mathematical Physics}\ }\textbf {\bibinfo {volume} {37}},\
  \bibinfo {pages} {4845–4866} (\bibinfo {year} {1996})}\BibitemShut
  {NoStop}%
\bibitem [{\citenamefont {Devoret}\ and\ \citenamefont
  {Grabert}(1992)}]{Grabert1992}%
  \BibitemOpen
  \bibfield  {author} {\bibinfo {author} {\bibfnamefont {M.~H.}\ \bibnamefont
  {Devoret}}\ and\ \bibinfo {author} {\bibfnamefont {H.}~\bibnamefont
  {Grabert}},\ }\href {http://dx.doi.org/10.1007/978-1-4757-2166-9} {\bibfield
  {journal} {\bibinfo  {journal} {NATO ASI Series}\ } (\bibinfo {year}
  {1992})}\BibitemShut {NoStop}%
\bibitem [{\citenamefont {Averin}\ and\ \citenamefont
  {Nazarov}(1992)}]{Averin1992}%
  \BibitemOpen
  \bibfield  {author} {\bibinfo {author} {\bibfnamefont {D.~V.}\ \bibnamefont
  {Averin}}\ and\ \bibinfo {author} {\bibfnamefont {Y.~V.}\ \bibnamefont
  {Nazarov}},\ }\href {\doibase 10.1103/physrevlett.69.1993} {\bibfield
  {journal} {\bibinfo  {journal} {Phys. Rev. Lett.}\ }\textbf {\bibinfo
  {volume} {69}},\ \bibinfo {pages} {1993–1996} (\bibinfo {year}
  {1992})}\BibitemShut {NoStop}%
\bibitem [{\citenamefont {Beenakker}\ \emph {et~al.}(2003)\citenamefont
  {Beenakker}, \citenamefont {Kindermann},\ and\ \citenamefont
  {Nazarov}}]{Beenakker2003}%
  \BibitemOpen
  \bibfield  {author} {\bibinfo {author} {\bibfnamefont {C.~W.~J.}\
  \bibnamefont {Beenakker}}, \bibinfo {author} {\bibfnamefont {M.}~\bibnamefont
  {Kindermann}}, \ and\ \bibinfo {author} {\bibfnamefont {Y.~V.}\ \bibnamefont
  {Nazarov}},\ }\href {\doibase 10.1103/PhysRevLett.90.176802} {\bibfield
  {journal} {\bibinfo  {journal} {Phys. Rev. Lett.}\ }\textbf {\bibinfo
  {volume} {90}},\ \bibinfo {pages} {176802} (\bibinfo {year}
  {2003})}\BibitemShut {NoStop}%
\bibitem [{\citenamefont {Reulet}\ \emph {et~al.}(2003)\citenamefont {Reulet},
  \citenamefont {Senzier},\ and\ \citenamefont {Prober}}]{Reulet2003}%
  \BibitemOpen
  \bibfield  {author} {\bibinfo {author} {\bibfnamefont {B.}~\bibnamefont
  {Reulet}}, \bibinfo {author} {\bibfnamefont {J.}~\bibnamefont {Senzier}}, \
  and\ \bibinfo {author} {\bibfnamefont {D.~E.}\ \bibnamefont {Prober}},\
  }\href {\doibase 10.1103/PhysRevLett.91.196601} {\bibfield  {journal}
  {\bibinfo  {journal} {Phys. Rev. Lett.}\ }\textbf {\bibinfo {volume} {91}},\
  \bibinfo {pages} {196601} (\bibinfo {year} {2003})}\BibitemShut {NoStop}%
\bibitem [{\citenamefont {Reulet}(2005)}]{Reulet2005}%
  \BibitemOpen
  \bibfield  {author} {\bibinfo {author} {\bibfnamefont {B.}~\bibnamefont
  {Reulet}},\ }\href {\doibase 10.1016/s0924-8099(05)80048-4} {\bibfield
  {journal} {\bibinfo  {journal} {Les Houches}\ ,\ \bibinfo {pages}
  {361–382}} (\bibinfo {year} {2005})}\BibitemShut {NoStop}%
\bibitem [{\citenamefont {Devoret}\ \emph {et~al.}(1990)\citenamefont
  {Devoret}, \citenamefont {Esteve}, \citenamefont {Grabert}, \citenamefont
  {Ingold}, \citenamefont {Pothier},\ and\ \citenamefont
  {Urbina}}]{Devoret1990}%
  \BibitemOpen
  \bibfield  {author} {\bibinfo {author} {\bibfnamefont {M.~H.}\ \bibnamefont
  {Devoret}}, \bibinfo {author} {\bibfnamefont {D.}~\bibnamefont {Esteve}},
  \bibinfo {author} {\bibfnamefont {H.}~\bibnamefont {Grabert}}, \bibinfo
  {author} {\bibfnamefont {G.-L.}\ \bibnamefont {Ingold}}, \bibinfo {author}
  {\bibfnamefont {H.}~\bibnamefont {Pothier}}, \ and\ \bibinfo {author}
  {\bibfnamefont {C.}~\bibnamefont {Urbina}},\ }\href {\doibase
  10.1103/physrevlett.64.1824} {\bibfield  {journal} {\bibinfo  {journal}
  {Phys. Rev. Lett.}\ }\textbf {\bibinfo {volume} {64}},\ \bibinfo {pages}
  {1824–1827} (\bibinfo {year} {1990})}\BibitemShut {NoStop}%
\bibitem [{\citenamefont {Ingold}\ and\ \citenamefont
  {Nazarov}(1992)}]{Ingold1992}%
  \BibitemOpen
  \bibfield  {author} {\bibinfo {author} {\bibfnamefont {G.-L.}\ \bibnamefont
  {Ingold}}\ and\ \bibinfo {author} {\bibfnamefont {Y.~V.}\ \bibnamefont
  {Nazarov}},\ }\href {http://dx.doi.org/10.1007/978-1-4757-2166-9_2}
  {\bibfield  {journal} {\bibinfo  {journal} {Single Charge Tunneling}\ ,\
  \bibinfo {pages} {Plenum Press, New York}} (\bibinfo {year}
  {1992})}\BibitemShut {NoStop}%
\bibitem [{\citenamefont {Delsing}\ \emph {et~al.}(1989)\citenamefont
  {Delsing}, \citenamefont {Likharev}, \citenamefont {Kuzmin},\ and\
  \citenamefont {Claeson}}]{Delsing1989}%
  \BibitemOpen
  \bibfield  {author} {\bibinfo {author} {\bibfnamefont {P.}~\bibnamefont
  {Delsing}}, \bibinfo {author} {\bibfnamefont {K.~K.}\ \bibnamefont
  {Likharev}}, \bibinfo {author} {\bibfnamefont {L.~S.}\ \bibnamefont
  {Kuzmin}}, \ and\ \bibinfo {author} {\bibfnamefont {T.}~\bibnamefont
  {Claeson}},\ }\href {\doibase 10.1103/physrevlett.63.1180} {\bibfield
  {journal} {\bibinfo  {journal} {Phys. Rev. Lett.}\ }\textbf {\bibinfo
  {volume} {63}},\ \bibinfo {pages} {1180–1183} (\bibinfo {year}
  {1989})}\BibitemShut {NoStop}%
\bibitem [{\citenamefont {Geerligs}\ \emph {et~al.}(1989)\citenamefont
  {Geerligs}, \citenamefont {Peters}, \citenamefont {de~Groot}, \citenamefont
  {Verbruggen},\ and\ \citenamefont {Mooij}}]{Geerligs1989}%
  \BibitemOpen
  \bibfield  {author} {\bibinfo {author} {\bibfnamefont {L.~J.}\ \bibnamefont
  {Geerligs}}, \bibinfo {author} {\bibfnamefont {M.}~\bibnamefont {Peters}},
  \bibinfo {author} {\bibfnamefont {L.~E.~M.}\ \bibnamefont {de~Groot}},
  \bibinfo {author} {\bibfnamefont {A.}~\bibnamefont {Verbruggen}}, \ and\
  \bibinfo {author} {\bibfnamefont {J.~E.}\ \bibnamefont {Mooij}},\ }\href
  {\doibase 10.1103/PhysRevLett.63.326} {\bibfield  {journal} {\bibinfo
  {journal} {Phys. Rev. Lett.}\ }\textbf {\bibinfo {volume} {63}},\ \bibinfo
  {pages} {326} (\bibinfo {year} {1989})}\BibitemShut {NoStop}%
\bibitem [{\citenamefont {Cleland}\ \emph {et~al.}(1992)\citenamefont
  {Cleland}, \citenamefont {Schmidt},\ and\ \citenamefont
  {Clarke}}]{Cleland1992}%
  \BibitemOpen
  \bibfield  {author} {\bibinfo {author} {\bibfnamefont {A.~N.}\ \bibnamefont
  {Cleland}}, \bibinfo {author} {\bibfnamefont {J.~M.}\ \bibnamefont
  {Schmidt}}, \ and\ \bibinfo {author} {\bibfnamefont {J.}~\bibnamefont
  {Clarke}},\ }\href {\doibase 10.1103/PhysRevB.45.2950} {\bibfield  {journal}
  {\bibinfo  {journal} {Phys. Rev. B}\ }\textbf {\bibinfo {volume} {45}},\
  \bibinfo {pages} {2950} (\bibinfo {year} {1992})}\BibitemShut {NoStop}%
\bibitem [{\citenamefont {Holst}\ \emph {et~al.}(1994)\citenamefont {Holst},
  \citenamefont {Esteve}, \citenamefont {Urbina},\ and\ \citenamefont
  {Devoret}}]{Holst1994}%
  \BibitemOpen
  \bibfield  {author} {\bibinfo {author} {\bibfnamefont {T.}~\bibnamefont
  {Holst}}, \bibinfo {author} {\bibfnamefont {D.}~\bibnamefont {Esteve}},
  \bibinfo {author} {\bibfnamefont {C.}~\bibnamefont {Urbina}}, \ and\ \bibinfo
  {author} {\bibfnamefont {M.~H.}\ \bibnamefont {Devoret}},\ }\href {\doibase
  10.1103/physrevlett.73.3455} {\bibfield  {journal} {\bibinfo  {journal}
  {Phys. Rev. Lett.}\ }\textbf {\bibinfo {volume} {73}},\ \bibinfo {pages}
  {3455–3458} (\bibinfo {year} {1994})}\BibitemShut {NoStop}%
\bibitem [{\citenamefont {Pierre}\ \emph {et~al.}(2001)\citenamefont {Pierre},
  \citenamefont {Pothier}, \citenamefont {Joyez}, \citenamefont {Birge},
  \citenamefont {Esteve},\ and\ \citenamefont {Devoret}}]{Pierre2001}%
  \BibitemOpen
  \bibfield  {author} {\bibinfo {author} {\bibfnamefont {F.}~\bibnamefont
  {Pierre}}, \bibinfo {author} {\bibfnamefont {H.}~\bibnamefont {Pothier}},
  \bibinfo {author} {\bibfnamefont {P.}~\bibnamefont {Joyez}}, \bibinfo
  {author} {\bibfnamefont {N.~O.}\ \bibnamefont {Birge}}, \bibinfo {author}
  {\bibfnamefont {D.}~\bibnamefont {Esteve}}, \ and\ \bibinfo {author}
  {\bibfnamefont {M.~H.}\ \bibnamefont {Devoret}},\ }\href {\doibase
  10.1103/physrevlett.86.1590} {\bibfield  {journal} {\bibinfo  {journal}
  {Phys. Rev. Lett.}\ }\textbf {\bibinfo {volume} {86}},\ \bibinfo {pages}
  {1590–1593} (\bibinfo {year} {2001})}\BibitemShut {NoStop}%
\bibitem [{\citenamefont {Safi}\ and\ \citenamefont
  {Sukhorukov}(2010)}]{Safi2010}%
  \BibitemOpen
  \bibfield  {author} {\bibinfo {author} {\bibfnamefont {I.}~\bibnamefont
  {Safi}}\ and\ \bibinfo {author} {\bibfnamefont {E.~V.}\ \bibnamefont
  {Sukhorukov}},\ }\href {http://stacks.iop.org/0295-5075/91/i=6/a=67008}
  {\bibfield  {journal} {\bibinfo  {journal} {EuroPhys. Lett.}\ }\textbf
  {\bibinfo {volume} {91}} (\bibinfo {year} {2010})}\BibitemShut {NoStop}%
\bibitem [{\citenamefont {Grabert}(2015)}]{Grabert2015}%
  \BibitemOpen
  \bibfield  {author} {\bibinfo {author} {\bibfnamefont {H.}~\bibnamefont
  {Grabert}},\ }\href {\doibase 10.1103/PhysRevB.92.245433} {\bibfield
  {journal} {\bibinfo  {journal} {Phys. Rev. B}\ }\textbf {\bibinfo {volume}
  {92}},\ \bibinfo {pages} {245433} (\bibinfo {year} {2015})}\BibitemShut
  {NoStop}%
\bibitem [{\citenamefont {Dolan}(1977)}]{Dolan1977}%
  \BibitemOpen
  \bibfield  {author} {\bibinfo {author} {\bibfnamefont {G.~J.}\ \bibnamefont
  {Dolan}},\ }\href {\doibase 10.1063/1.89690} {\bibfield  {journal} {\bibinfo
  {journal} {App. Phys. Lett.}\ }\textbf {\bibinfo {volume} {31}},\ \bibinfo
  {pages} {337–339} (\bibinfo {year} {1977})}\BibitemShut {NoStop}%
\bibitem [{\citenamefont {Tien}\ and\ \citenamefont {Gordon}(1963)}]{Tien1963}%
  \BibitemOpen
  \bibfield  {author} {\bibinfo {author} {\bibfnamefont {P.~K.}\ \bibnamefont
  {Tien}}\ and\ \bibinfo {author} {\bibfnamefont {J.~P.}\ \bibnamefont
  {Gordon}},\ }\href {\doibase 10.1103/physrev.129.647} {\bibfield  {journal}
  {\bibinfo  {journal} {Phys Rev.}\ }\textbf {\bibinfo {volume} {129}},\
  \bibinfo {pages} {647–651} (\bibinfo {year} {1963})}\BibitemShut {NoStop}%
\bibitem [{\citenamefont {Schoelkopf}\ \emph {et~al.}(1998)\citenamefont
  {Schoelkopf}, \citenamefont {Kozhevnikov}, \citenamefont {Prober},\ and\
  \citenamefont {Rooks}}]{Schoelkopf1998}%
  \BibitemOpen
  \bibfield  {author} {\bibinfo {author} {\bibfnamefont {R.~J.}\ \bibnamefont
  {Schoelkopf}}, \bibinfo {author} {\bibfnamefont {A.~A.}\ \bibnamefont
  {Kozhevnikov}}, \bibinfo {author} {\bibfnamefont {D.~E.}\ \bibnamefont
  {Prober}}, \ and\ \bibinfo {author} {\bibfnamefont {M.~J.}\ \bibnamefont
  {Rooks}},\ }\href {\doibase 10.1103/physrevlett.80.2437} {\bibfield
  {journal} {\bibinfo  {journal} {Phys. Rev. Lett.}\ }\textbf {\bibinfo
  {volume} {80}},\ \bibinfo {pages} {2437–2440} (\bibinfo {year}
  {1998})}\BibitemShut {NoStop}%
\bibitem [{\citenamefont {Gabelli}\ and\ \citenamefont
  {Reulet}(2008)}]{Gabelli2008}%
  \BibitemOpen
  \bibfield  {author} {\bibinfo {author} {\bibfnamefont {J.}~\bibnamefont
  {Gabelli}}\ and\ \bibinfo {author} {\bibfnamefont {B.}~\bibnamefont
  {Reulet}},\ }\href {\doibase 10.1103/PhysRevLett.100.026601} {\bibfield
  {journal} {\bibinfo  {journal} {Phys. Rev. Lett.}\ }\textbf {\bibinfo
  {volume} {100}},\ \bibinfo {pages} {026601} (\bibinfo {year}
  {2008})}\BibitemShut {NoStop}%
\bibitem [{\citenamefont {Zakka-Bajjani}\ \emph {et~al.}(2007)\citenamefont
  {Zakka-Bajjani}, \citenamefont {S\'egala}, \citenamefont {Portier},
  \citenamefont {Roche}, \citenamefont {Glattli}, \citenamefont {Cavanna},\
  and\ \citenamefont {Jin}}]{Zakka2007}%
  \BibitemOpen
  \bibfield  {author} {\bibinfo {author} {\bibfnamefont {E.}~\bibnamefont
  {Zakka-Bajjani}}, \bibinfo {author} {\bibfnamefont {J.}~\bibnamefont
  {S\'egala}}, \bibinfo {author} {\bibfnamefont {F.}~\bibnamefont {Portier}},
  \bibinfo {author} {\bibfnamefont {P.}~\bibnamefont {Roche}}, \bibinfo
  {author} {\bibfnamefont {D.~C.}\ \bibnamefont {Glattli}}, \bibinfo {author}
  {\bibfnamefont {A.}~\bibnamefont {Cavanna}}, \ and\ \bibinfo {author}
  {\bibfnamefont {Y.}~\bibnamefont {Jin}},\ }\href {\doibase
  10.1103/PhysRevLett.99.236803} {\bibfield  {journal} {\bibinfo  {journal}
  {Phys. Rev. Lett.}\ }\textbf {\bibinfo {volume} {99}},\ \bibinfo {pages}
  {236803} (\bibinfo {year} {2007})}\BibitemShut {NoStop}%
\bibitem [{\citenamefont {Shytov}(2005)}]{Shytov2005}%
  \BibitemOpen
  \bibfield  {author} {\bibinfo {author} {\bibfnamefont {A.~V.}\ \bibnamefont
  {Shytov}},\ }\href {\doibase 10.1103/PhysRevB.71.085301} {\bibfield
  {journal} {\bibinfo  {journal} {Phys. Rev. B}\ }\textbf {\bibinfo {volume}
  {71}},\ \bibinfo {pages} {085301} (\bibinfo {year} {2005})}\BibitemShut
  {NoStop}%
\bibitem [{\citenamefont {Bagrets}\ and\ \citenamefont
  {Pistolesi}(2007)}]{Bagrets2007}%
  \BibitemOpen
  \bibfield  {author} {\bibinfo {author} {\bibfnamefont {D.}~\bibnamefont
  {Bagrets}}\ and\ \bibinfo {author} {\bibfnamefont {F.}~\bibnamefont
  {Pistolesi}},\ }\href {\doibase 10.1103/PhysRevB.75.165315} {\bibfield
  {journal} {\bibinfo  {journal} {Phys. Rev. B}\ }\textbf {\bibinfo {volume}
  {75}},\ \bibinfo {pages} {165315} (\bibinfo {year} {2007})}\BibitemShut
  {NoStop}%
\bibitem [{\citenamefont {Matveev}\ \emph {et~al.}(1993)\citenamefont
  {Matveev}, \citenamefont {Yue},\ and\ \citenamefont {Glazman}}]{Matveev1993}%
  \BibitemOpen
  \bibfield  {author} {\bibinfo {author} {\bibfnamefont {K.~A.}\ \bibnamefont
  {Matveev}}, \bibinfo {author} {\bibfnamefont {D.}~\bibnamefont {Yue}}, \ and\
  \bibinfo {author} {\bibfnamefont {L.~I.}\ \bibnamefont {Glazman}},\ }\href
  {\doibase 10.1103/physrevlett.71.3351} {\bibfield  {journal} {\bibinfo
  {journal} {Phys. Rev. Lett.}\ }\textbf {\bibinfo {volume} {71}},\ \bibinfo
  {pages} {3351–3354} (\bibinfo {year} {1993})}\BibitemShut {NoStop}%
\bibitem [{\citenamefont {Kindermann}\ and\ \citenamefont
  {Nazarov}(2003)}]{Kindermann2003}%
  \BibitemOpen
  \bibfield  {author} {\bibinfo {author} {\bibfnamefont {M.}~\bibnamefont
  {Kindermann}}\ and\ \bibinfo {author} {\bibfnamefont {Y.~V.}\ \bibnamefont
  {Nazarov}},\ }\href {\doibase 10.1103/PhysRevLett.91.136802} {\bibfield
  {journal} {\bibinfo  {journal} {Phys. Rev. Lett.}\ }\textbf {\bibinfo
  {volume} {91}},\ \bibinfo {pages} {136802} (\bibinfo {year}
  {2003})}\BibitemShut {NoStop}%
\bibitem [{\citenamefont {Gasse}\ \emph {et~al.}(2013)\citenamefont {Gasse},
  \citenamefont {Spietz}, \citenamefont {Lupien},\ and\ \citenamefont
  {Reulet}}]{Gasse2013}%
  \BibitemOpen
  \bibfield  {author} {\bibinfo {author} {\bibfnamefont {G.}~\bibnamefont
  {Gasse}}, \bibinfo {author} {\bibfnamefont {L.}~\bibnamefont {Spietz}},
  \bibinfo {author} {\bibfnamefont {C.}~\bibnamefont {Lupien}}, \ and\ \bibinfo
  {author} {\bibfnamefont {B.}~\bibnamefont {Reulet}},\ }\href {\doibase
  10.1103/PhysRevB.88.241402} {\bibfield  {journal} {\bibinfo  {journal} {Phys.
  Rev. B}\ }\textbf {\bibinfo {volume} {88}},\ \bibinfo {pages} {241402}
  (\bibinfo {year} {2013})}\BibitemShut {NoStop}%
\bibitem [{\citenamefont {Altimiras}\ \emph {et~al.}(2014)\citenamefont
  {Altimiras}, \citenamefont {Parlavecchio}, \citenamefont {Joyez},
  \citenamefont {Vion}, \citenamefont {Roche}, \citenamefont {Esteve},\ and\
  \citenamefont {Portier}}]{Altimiras2014}%
  \BibitemOpen
  \bibfield  {author} {\bibinfo {author} {\bibfnamefont {C.}~\bibnamefont
  {Altimiras}}, \bibinfo {author} {\bibfnamefont {O.}~\bibnamefont
  {Parlavecchio}}, \bibinfo {author} {\bibfnamefont {P.}~\bibnamefont {Joyez}},
  \bibinfo {author} {\bibfnamefont {D.}~\bibnamefont {Vion}}, \bibinfo {author}
  {\bibfnamefont {P.}~\bibnamefont {Roche}}, \bibinfo {author} {\bibfnamefont
  {D.}~\bibnamefont {Esteve}}, \ and\ \bibinfo {author} {\bibfnamefont
  {F.}~\bibnamefont {Portier}},\ }\href
  {http://dx.doi.org/10.1103/PhysRevLett.112.236803} {\bibfield  {journal}
  {\bibinfo  {journal} {Phys. Rev. Lett.}\ }\textbf {\bibinfo {volume} {112}}
  (\bibinfo {year} {2014})}\BibitemShut {NoStop}%
\bibitem [{\citenamefont {Lesovik}\ and\ \citenamefont
  {Levitov}(1994)}]{Levitov1994}%
  \BibitemOpen
  \bibfield  {author} {\bibinfo {author} {\bibfnamefont {G.~B.}\ \bibnamefont
  {Lesovik}}\ and\ \bibinfo {author} {\bibfnamefont {L.~S.}\ \bibnamefont
  {Levitov}},\ }\href {\doibase 10.1103/PhysRevLett.72.538} {\bibfield
  {journal} {\bibinfo  {journal} {Phys. Rev. Lett.}\ }\textbf {\bibinfo
  {volume} {72}},\ \bibinfo {pages} {538} (\bibinfo {year} {1994})}\BibitemShut
  {NoStop}%
\bibitem [{Note1()}]{Note1}%
  \BibitemOpen
  \bibinfo {note} {Similar extra steps have been reported in \cite
  {Altimiras2014} and attributed to the noise being measured given by
  non-symmetrized current-current correlator. This discussion is irrelevant
  here since our measurement corresponds to zero frequency noise, for which
  commutativity of current operators is not an issue.}\BibitemShut {Stop}%
\bibitem [{\citenamefont {Parlavecchio}\ \emph {et~al.}(2015)\citenamefont
  {Parlavecchio}, \citenamefont {Altimiras}, \citenamefont {Souquet},
  \citenamefont {Simon}, \citenamefont {Safi}, \citenamefont {Joyez},
  \citenamefont {Vion}, \citenamefont {Roche}, \citenamefont {Esteve},\ and\
  \citenamefont {Portier}}]{Parlavecchio2015}%
  \BibitemOpen
  \bibfield  {author} {\bibinfo {author} {\bibfnamefont {O.}~\bibnamefont
  {Parlavecchio}}, \bibinfo {author} {\bibfnamefont {C.}~\bibnamefont
  {Altimiras}}, \bibinfo {author} {\bibfnamefont {J.-R.}\ \bibnamefont
  {Souquet}}, \bibinfo {author} {\bibfnamefont {P.}~\bibnamefont {Simon}},
  \bibinfo {author} {\bibfnamefont {I.}~\bibnamefont {Safi}}, \bibinfo {author}
  {\bibfnamefont {P.}~\bibnamefont {Joyez}}, \bibinfo {author} {\bibfnamefont
  {D.}~\bibnamefont {Vion}}, \bibinfo {author} {\bibfnamefont {P.}~\bibnamefont
  {Roche}}, \bibinfo {author} {\bibfnamefont {D.}~\bibnamefont {Esteve}}, \
  and\ \bibinfo {author} {\bibfnamefont {F.}~\bibnamefont {Portier}},\ }\href
  {\doibase 10.1103/PhysRevLett.114.126801} {\bibfield  {journal} {\bibinfo
  {journal} {Phys. Rev. Lett.}\ }\textbf {\bibinfo {volume} {114}},\ \bibinfo
  {pages} {126801} (\bibinfo {year} {2015})}\BibitemShut {NoStop}%
\bibitem [{\citenamefont {Joyez}\ \emph {et~al.}(1998)\citenamefont {Joyez},
  \citenamefont {Esteve},\ and\ \citenamefont {Devoret}}]{Joyez1998}%
  \BibitemOpen
  \bibfield  {author} {\bibinfo {author} {\bibfnamefont {P.}~\bibnamefont
  {Joyez}}, \bibinfo {author} {\bibfnamefont {D.}~\bibnamefont {Esteve}}, \
  and\ \bibinfo {author} {\bibfnamefont {M.~H.}\ \bibnamefont {Devoret}},\
  }\href {\doibase 10.1103/physrevlett.80.1956} {\bibfield  {journal} {\bibinfo
   {journal} {Phys. Rev. Lett.}\ }\textbf {\bibinfo {volume} {80}},\ \bibinfo
  {pages} {1956–1959} (\bibinfo {year} {1998})}\BibitemShut {NoStop}%
\bibitem [{\citenamefont {Altimiras}\ \emph {et~al.}(2016)\citenamefont
  {Altimiras}, \citenamefont {Portier},\ and\ \citenamefont
  {Joyez}}]{Altimiras2016}%
  \BibitemOpen
  \bibfield  {author} {\bibinfo {author} {\bibfnamefont {C.}~\bibnamefont
  {Altimiras}}, \bibinfo {author} {\bibfnamefont {F.}~\bibnamefont {Portier}},
  \ and\ \bibinfo {author} {\bibfnamefont {P.}~\bibnamefont {Joyez}},\ }\href
  {\doibase 10.1103/PhysRevX.6.031002} {\bibfield  {journal} {\bibinfo
  {journal} {Phys. Rev. X}\ }\textbf {\bibinfo {volume} {6}},\ \bibinfo {pages}
  {031002} (\bibinfo {year} {2016})}\BibitemShut {NoStop}%
\bibitem [{Note2()}]{Note2}%
  \BibitemOpen
  \bibinfo {note} {Indeed, we note that our photon-assisted transport data are
  well fitted by taking replicas of the data at $V_{ac}=0$ and the lowest
  temperature, whereas noise data at high excitation show an increased
  effective temperature.}\BibitemShut {Stop}%
\bibitem [{Note3()}]{Note3}%
  \BibitemOpen
  \bibinfo {note} {Such effects have been discussed in similar samples where
  the amplitude of quantum oscillations could not be explained by the ac
  voltage excitation deduced from photon-assisted noise measurements \cite
  {Gasse2013}.}\BibitemShut {Stop}%
\bibitem [{\citenamefont {Souquet}\ \emph {et~al.}(2014)\citenamefont
  {Souquet}, \citenamefont {Woolley}, \citenamefont {Gabelli}, \citenamefont
  {Simon},\ and\ \citenamefont {Clerk}}]{Souquet2014}%
  \BibitemOpen
  \bibfield  {author} {\bibinfo {author} {\bibfnamefont {J.~R.}\ \bibnamefont
  {Souquet}}, \bibinfo {author} {\bibfnamefont {M.~J.}\ \bibnamefont
  {Woolley}}, \bibinfo {author} {\bibfnamefont {J.}~\bibnamefont {Gabelli}},
  \bibinfo {author} {\bibfnamefont {P.}~\bibnamefont {Simon}}, \ and\ \bibinfo
  {author} {\bibfnamefont {A.~A.}\ \bibnamefont {Clerk}},\ }\href {\doibase
  10.1038/ncomms6562} {\bibfield  {journal} {\bibinfo  {journal} {Nat. Comm.}\
  }\textbf {\bibinfo {volume} {5}},\ \bibinfo {pages} {5562} (\bibinfo {year}
  {2014})}\BibitemShut {NoStop}%
\end{thebibliography}%

\end{document}